\documentclass[12pt]{article}
\usepackage{latexsym}
\usepackage{amsfonts}
\usepackage{amssymb}
\usepackage{amsmath}
\usepackage{amsthm}

\newcommand{\cC}{{\mathcal C}}
\newcommand{\cP}{{\mathcal P}}
\newcommand{\bP}{\bar{\mathcal P}}
\newcommand{\bC}{{\bar{\mathcal C}}}
\newcommand{\myth}{\mathbf{\Theta}}
\newcommand{\myths}{\mathbf{\theta}}

\numberwithin{equation}{section}

\newcommand{\bref}[1]{\textbf{\ref{#1}}}

\def\H{\mathcal H}

\def\ih{i\hbar\,}

\newcommand{\pb}[2]{\displaystyle{\left\{{}#1{},{}#2{}\right\}}}
\newcommand{\ab}[2]{\big(#1,\,#2\big)}

\newcommand{\commut}[2]{[#1,#2]}

\def\half{{\frac{1}{2}}}
\newcommand{\gh}[1]{{\rm gh}(#1)}
\newcommand{\p}[1]{\varepsilon{(#1)}}
\newcommand{\ip}[1]{\varepsilon_{#1}}
\newcommand{\hepth}[1]{\texttt{#1}}

\begin{document}
\addtolength{\baselineskip}{2.3pt}

{\hfill{\lowercase{\tt hep-th/0301085}}}

{\hfill{ULB-TH/03-10}}\\[12pt]

\begin{centering}
{\Large \textbf{BRST-anti-BRST Symmetric Conversion of Second-Class
    Constraints}}\\

\vspace{1cm}

{\large{I.A.~Batalin$~^a$ and M.A.~Grigoriev$~^{a,b}$}}\\

\vspace{0.5cm}

$~^{a}$ Lebedev Physics
  Institute, RAS, Leninskiy 53, Moscow 119991, Russia\\
$~^b$ Physique Th\'eorique et Math\'ematique,
 Universit\'e Libre
 de Bruxelles,  C.P. 231, B--1050, Bruxelles, Belgium \\
\end{centering}

\vspace{0.6cm}

\begin{abstract}
  A general method of the BRST--anti-BRST symmetric conversion of
  second-class constraints is presented.  It yields a pair of
  commuting and nilpotent BRST-type charges that can be naturally
  regarded as BRST and anti-BRST ones.  Interchanging the BRST and
  anti-BRST generators corresponds to a symmetry between the original
  second-class constraints and the conversion variables, which
  enter the formalism on equal footing.
\end{abstract}

\vspace{0.7cm}

\thispagestyle{empty}

\section{Introduction}
For first-class constraints, the well-known BFV--BRST quantization
method provides an adequate description of general constrained systems
at the classical and quantum levels~\cite{BFV} (for review see
\cite{HT}).  But for second-class
systems, a proper counterpart of the BFV--BRST method is not yet
known.  Most of the well-established approaches to second-class
systems are based on the construction of an effective first-class
system that is equivalent to the original second-class
system~\cite{FSh,BF,BF87, Split,BMgauge,BLM-Proj}.  Among these methods, the
most developed one is the so-called \textit{conversion
  method}~\cite{FSh,BF,BF87,BT,BFF,BGL}.  In this approach, one introduces
additional variables $\phi$, called \textit{conversion variables}, and
extends second-class constraints by $\phi$-dependent terms such that
the extended constraints become first-class.

The conversion method has proved a powerful tool for description and
quantization of second-class systems.  But the general conversion is
too ambiguous, and it seems natural to reduce this ambiguity by
imposing additional conditions on the conversion procedure.  A
well-known restriction consists in requiring the effective gauge
algebra to be of a fixed rank; a common choice is to take a $0$-rank
(Abelian) effective gauge algebra.  This \textit{Abelian} conversion
is well studied; in particular, the existence theorem and a
constructive procedure are known for the effective
constraints~\cite{BT}.

In this paper, we take another route and reduce the ambiguity by
requiring the conversion procedure to respect extra symmetries.
Namely, we treat the original second-class constraints and conversion
variables on equal footing.  Technically, this implies that the
conversion variables appear as constraints $\bar\theta^\alpha$ dual to
the original second-class constraints $\theta_\alpha$.  We then
proceed with the conversion keeping the symmetry between
$\theta_\alpha$ and $\bar\theta^\alpha$ explicit.

The conversion is carried out in the framework of the BFV--BRST
approach.  At the level of the effective gauge system (converted
system), the symmetry between $\theta^\alpha$ and $\bar\theta^\alpha$
results in the BRST--anti-BRST invariance.  The BRST and anti-BRST charges
are associated to the appropriately
extended original constraints $\theta_\alpha$ and the extended dual
constraints $\bar\theta^\alpha$ respectively.

The paper is organized as follows: in Sec.~\bref{sec:basics}, we
develop basics of our approach to conversion and show that the
resulting gauge system is naturally described by a pair of nilpotent
and commuting generators identified as the BRST and anti-BRST
generators.  In Sec.~\bref{sec:QA}, generating equations of the
BRST--anti-BRST symmetric conversion are reformulated as a certain
type of master equations with respect to the~\textit{quantum
  antibracket}~\cite{BMQA}.  A more restricted version of the
formalism is considered in Sec.~\bref{sec:special}.  In this more
special approach, generating operators satisfy stronger conditions.
These conditions are also shown to allow the existence of a conversion
procedure for arbitrary original constraints $\theta_\alpha$.  In
Sec.~\bref{sec:Hamiltonian}, we consider construction of a unitarizing
Hamiltonian in the BRST--anti-BRST symmetric formulation.

\section{Basics}\label{sec:basics}
Let a second-class constrained system be determined by constraints
$\theta^\alpha\,,\alpha=1,\ldots\,,2N$.  Their Dirac matrix is given
by
\begin{equation}
\ih  \Delta^0_{\alpha\beta}=\commut{\theta_\alpha}{\theta_\beta}
\end{equation}
and is assumed to be invertible. To each constraint $\theta^\alpha$ we
associate its dual $\bar\theta^\alpha$. We choose
$\p{\bar\theta^\alpha}=\p{\theta_\alpha}$ where $\p{f}$ denotes
Grassmann parity of $f$. Dual constraints plays a role of extra
degrees of freedom and therefore are defined on the appropriate
extension of the original phase space.  A basic example is provided by
taking $\bar\theta^\alpha=e_\beta^\alpha \phi^\beta$ where
$\phi^\alpha$ are conversion variables and $e^\alpha_\beta$ depend on
phase space variables only.

In what follows we prefer to work in terms of generating functions (operators)
and not in terms of constraints. To this end let us introduce ghost variables
$\cC^\alpha$ and $\bP_\alpha$ associated either to
constraints $\theta_\alpha$ or $\bar \theta^\alpha$. One assignes the 
following
Grassmann parities to the
ghost variables:
\begin{equation}
\p{\cC^\alpha}=\p{\bP_\alpha}=\p{\theta_\alpha}+1=\p{\bar\theta^\alpha}+1\,, 
\qquad
\alpha=1,\ldots,2N\,.
\end{equation}
We also choose standard commutation relations and ghost number gradings
for variables $\cC^\alpha$ and $\bP_\alpha$
\begin{equation}
\commut{\cC^\alpha}{\bP_\beta}=\ih \delta^\alpha_\beta\,,\qquad
\gh{\cC^\alpha}=1\,, \quad \gh{\bP_\alpha}= -1\,.
\end{equation}
The ghost number operator is then assumed to have the following
form:
\begin{equation}
G=\half(\cC^\alpha\bP_\alpha(-1)^{\ip{\alpha}}-\bP_\alpha\cC^\alpha)\,.
\end{equation}
Generating operators encoding constraints $\theta_\alpha$
and $\bar\theta^\alpha$ are denoted by $\myth$ and $\bar\myth$
respectively; their expansions with respect to ghost variables start as
\begin{equation}
\myth=\cC^\alpha \theta_\alpha+\ldots\,,\qquad
\bar\myth=\bar\theta^\alpha \bP_\alpha (-1)^{\ip{\alpha}}+\ldots\,,
\end{equation}
and are subjected to the following ghost number prescriptions:
\begin{equation}
  \commut{G}{\myth}=\ih \myth\,,\qquad
  \commut{G}{\bar\myth}=-\ih\bar\myth\,.
\end{equation}

We are interested in constructing first-class constrained (gauge) system
equivalent to the original second-class one and entering constraints $\theta$
and $\bar\theta$ in a symmetric way. To this end we are looking for a pair
of generators $\Omega$ and $\bar\Omega$ satisfying
\begin{equation}
  \begin{gathered}
\label{eq:nilpotency}
\commut{\Omega}{\Omega}=0\,,\qquad
\commut{\bar\Omega}{\bar\Omega}=0\,,\\
\commut{\Omega}{\bar\Omega}=0\,.
  \end{gathered}
\end{equation}
$\Omega$ and $\bar\Omega$ are to be understood as appropriate
extensions of the original generating operators $\myth$ and
$\bar\myth$ respectively.

It is useful to introduce the following condensed notations:
\begin{equation}
  \begin{aligned}
 \myth^{a}&\,,\quad a=1,2\, \qquad
 \myth^1=\myth\,,\quad
\myth^2=\bar\myth\,,\\
 {\Omega^{a}}&\,,\quad a=1,2\, \qquad
 {\Omega^1}={\Omega}\,,\quad 
{\Omega^2}={\bar\Omega}\,.
\end{aligned}
\end{equation}
The ghost number assignments then take the form
\begin{equation}
\commut{G}{\myth^a}=\ih g^a_b \myth^b \,,
\qquad
\commut{G}{{\Omega^a}}=
\ih g^a_b {\Omega^b}\,,\label{eq:ghost}
\end{equation}
where the only nonvanishing components of $g^a_b$ are
$g^1_1=1$ and $g^2_2=-1$. We also introduce operator $\Delta^{ab}$
defined by
\begin{equation}
2\ih\Delta^{ab}=\commut{\myth^a}{\myth^b}\,.
\end{equation}
In terms of new notations conditions \eqref{eq:nilpotency}
read as
\begin{equation}
\label{eq:nilpotency-short}
  \commut{{\Omega^a}}{{\Omega^b}}=0\,.
\end{equation}

In order to find $\Omega^a$ satisfying~\eqref{eq:nilpotency-short}
consider the following anzatz for ${\Omega^a}$
\begin{equation}
{\Omega^a}=\myth^a+ (\ih)^{-1}\commut{Z}{\myth^a}\,,
\end{equation}
where an operator $Z$ has been introduced.
Equations~\eqref{eq:nilpotency-short} take then the form:
\begin{multline}
\label{eq:explicit}
\quad 2\, \commut{{\Omega^a}}{{\Omega^b}}
~=~
\commut{{\Omega^{\{a}}}{{\Omega^{b\}}}}
~=~
\,
\\
~=~
\commut{\myth^{\{a}}{\myth^{b\}}-(\ih)^{-2}\commut{Z}{\commut{Z}{\myth^{b\}}}}}
~+~
\\
~+~
4\commut{Z}{\Delta^{ab}}
~ + ~2\,(\ih)^{-1}
\commut{Z}{\commut{Z}{\Delta^{ab}}}~=~0\,. \quad
\end{multline}
It is natural to make this conditions satisfied by imposing the following
stronger conditions
\begin{equation}
\label{eq:cond1}
\begin{split}
\commut{Z}{\commut{Z}{\myth^a}}&=(\ih)^2\myth^a\,,\\
\half\commut{Z}{\commut{Z}{\Delta^{ab}}}&=-\ih\commut{Z}{\Delta^{ab}}\,,
\end{split}
\end{equation}
so that the second and third lines in~\eqref{eq:explicit} vanish separately.
It follows from~\eqref{eq:cond1} that $Z$ acts on
${\Omega^a}=\myth^a+ (\ih)^{-1}\commut{Z}{\myth^a}$ as a projector:
\begin{equation}
  \commut{Z}{{\Omega^c}}=i\hbar {\Omega^c}\,.
\end{equation}

Given $\myth^a$ and $Z$ such that \eqref{eq:cond1} holds one arrives at a pair
of commuting and nilpotent BRST-like charges ${\Omega}^a$. They can be
understood as BRST and anti-BRST generators. To explain this in more details
let us consider explicitly the lowest order equations coming
from~\eqref{eq:nilpotency}.
Introducing the lowest
order structure functions according to
\begin{equation}
\begin{aligned}
\label{eq:Omega-exp}
\Omega&={\Omega}^1=\cC^\alpha {T}_\alpha
+
\half \cC^\beta \cC^\alpha {U}^\gamma_{\alpha\beta}\bP_\gamma
(-1)^{\ip{\beta}+\ip{\gamma}}
\,+\,\ldots\,,
\\
\bar\Omega&=
{\Omega}^2
=
{\bar{T}}^\alpha \bP_\alpha(-1)^{\ip{\alpha}}
+
\half \cC^\gamma {\bar{U}}^{\alpha\beta}_\gamma
\bP_\beta \bP_\alpha(-1)^{\ip{\beta}}
\,+\,\ldots\,,
\end{aligned}
\end{equation}
one arrives at the following explicit form
of the lowest order equations:
\begin{gather}
\label{eq:1st-cond}
{\bar{T}}^\alpha T_\alpha =0\,,\\
\commut{T_\alpha}{T_\beta}=
i\hbar\, {U}^\gamma_{\alpha\beta}T_\gamma\,,
\qquad
\commut{{\bar{T}}^\alpha}{{\bar{T}}^\beta}=
i\hbar\,{\bar{T}}^\gamma {\bar{U}}_\gamma^{\alpha\beta}\,,\\
\commut{T_\alpha}{{\bar{T}}^\beta}
=
i\hbar\,({\bar{U}}^{\beta\gamma}_\alpha T_\gamma
+
{\bar{T}}^\gamma {U}^\beta_{\gamma\alpha})+O(\hbar^2)\,.
\end{gather}
These relations coincide with those familiar from the standard BRST-anti-BRST
symmetric formulation.

Let us also note that the algebra formed by $\Omega^a$ and $Z$
coincides with that in the $sp(2)$ symmetric formalism with
$Z$ being a counterpart of the ``new ghost number''
operator~\cite{BLT}.

\section{Formulation in terms of the quantum antibracket}\label{sec:QA}
Generating equations~\eqref{eq:cond1} can be interpreted as
a type of master equations formulated in terms of the so-called
\textit{quantum antibracket}. To show that consider the
following bracket operation:
\begin{equation}
\ab{f}{g}_Q=\half
\left(
\commut{f}{\commut{Q}{g}}+(-1)^{(\p{f}+\p{Q})(\p{g}+\p{Q})+\p{Q}}
\commut{g}{\commut{Q}{f}}
\right)\,.
\end{equation}
In the case where $\p{Q}=1$ this structure is know as the
quantum antibracket~\cite{BMQA}. In terms of this structure
equations~\eqref{eq:cond1} take the following form:
\begin{equation}
  \begin{split}
\label{eq:masterZ}
\ab{Z}{Z}_{\myth^a}&=-(\ih)^2\myth^a\,,\\
\half\ab{Z}{Z}_{\Delta^{ab}}&=-\ih\commut{\Delta^{ab}}{Z}\,.
\end{split}
\end{equation}

One can also reformulate these equations in terms of
a projection operator 
\begin{equation}
S^a_b=g^a_b G+ \delta^a_b Z\,,
\end{equation}
so that one can express $\Omega^a$ as follows
\begin{equation}
\Omega^a=(i\hbar)^{-1}\commut{S^a_b}{\myth^b}\,.
\end{equation}
Indeed, it follows from~\eqref{eq:masterZ} that $S^a_b$
satisfies
\begin{align}
\ab{S^{\{a}_{\{c}}{S^{b\}}_{d\}}}_{\myth^d}
&=i\hbar 
\left(
g^{\{a}_{\{c}g^{b\}}_{d\}}+ \delta^{\{a}_{\{c}\delta^{b\}}_{d\}}
\right)
\commut{\myth^e}{S^d_e}\,,
\\
\half\ab{S^{\{a}_{\{c}}{S^{b\}}_{d\}}}_{\Delta^{cd}}
&= i\hbar \,
g^{\{a}_{\{c}g^{b\}}_{d\}}\,
\commut{\Delta^{ed}}{S^c_e}\,.
\end{align}

\section{More special formulation} \label{sec:special}

As we have seen conditions~\eqref{eq:cond1} ensure nilpotency
and mutual commutativity of ${\Omega^a}$. One can either
consider $\myth^a$ and $Z$ as independent quantities to be
defined by~\eqref{eq:cond1} or try to reduce the number
of unknowns. The later option leads a priory to a more
restricted class of solutions to~\eqref{eq:cond1}.
We choose this option and express  $Z$ through $\myth^a$
by the following anzatz
\begin{equation}
\ih Z =\commut{\myth}{\bar\myth} = \ih g_{ab}\Delta^{ab}\,,\label{eq:Z}
\end{equation}
where $g_{ab}=\epsilon_{ac} g^c_b$ with $\epsilon_{ab}$ being antisymmetric 
and
$\epsilon_{12}=-1$. As a consequence operators $\myth,\bar\myth$,$\Delta^{ab}$
and $Z$ satisfy the following relations:
\begin{equation}
\commut{\Delta}{\bar\myth}=\commut{\myth}{Z}\,,\qquad
\commut{\bar\Delta}{\myth}=\commut{\bar\myth}{Z}\,,
\end{equation}
where notations
\begin{equation}
2 i\hbar\, \Delta=2 i\hbar\, \Delta^{11}=\commut{\myth}{\myth}\,, \qquad
2 i\hbar\, \bar\Delta=2 i\hbar\, \Delta^{22}=\commut{\bar\myth}{\bar\myth}\,,
\end{equation}
are introduced.

In the case where $Z$ is given by~\eqref{eq:Z} conditions~\eqref{eq:cond1}
can be satisfied by imposing the following stronger ones:
\begin{equation}
  \begin{gathered}
\label{eq:basic}
    \commut{Z}{\Delta}=0\,,\qquad \commut{Z}{\bar\Delta}=0\,,\\
\commut{\Delta}{\bar\Delta}=\ih G\,.
  \end{gathered}
\end{equation}
These are to be understood as equations on $\myth^a$
ensuring nilpotency and mutual commutativity of
${\Omega^a}=\myth^a+(i\hbar)^{-1}\commut{Z}{\myth^a}$.
It also follows from Eq.~\eqref{eq:basic} that
\begin{equation}
\commut{\myth^{\{a}}{\commut{\myth^{b\}}}{Z}}=0\,.
\end{equation}
This in turn implies that one can consider the following a little
bit more general expressions for $\Omega^a$
\begin{equation}
\Omega=\myth+\alpha(i\hbar)^{-1}\commut{Z}{\myth}\,, \qquad
\bar\Omega=\bar\myth+\bar\alpha(i\hbar)^{-1}\commut{Z}{\bar\myth}\,,
\end{equation}
which are nilpotent provided $\alpha=\bar\alpha=1$ or
$\alpha=\bar\alpha=-1$.

It is instructive to rewrite equations~\eqref{eq:basic}
in a condensed form directly in terms
of $\Delta^{ab}$:
\begin{equation}
\label{eq:Delta-comm}
  4\commut{\Delta^{ab}}{\Delta^{cd}}
=
i\hbar g^{\{a\{c } \epsilon^{b\}d\}} G\,, 
\end{equation}
where
\begin{equation}
g^{\{a\{c } \epsilon^{b\}d\}}=
g^{ac}\epsilon^{bd}+g^{bc}\epsilon^{ad}+
g^{ad}\epsilon^{bc}+g^{bd}\epsilon^{ac}
\end{equation}
and $g^{ab}=g^a_c \epsilon^{cb}$ with $\epsilon^{12}=1$. One can
then check that these equations do contain all relations~\eqref{eq:basic}.

Contracting equation~\eqref{eq:Delta-comm} with $g_{ab}=\epsilon_{ac}g^c_b$
and $\epsilon_{ab}$ one arrives at
\begin{equation}
  \epsilon_{bc}\commut{\Delta^{ab}}{\Delta^{cd}}=- i\hbar g^{ad} G\,
\end{equation}
and
\begin{equation}
  g_{bc}\commut{\Delta^{ab}}{\Delta^{cd}}= i\hbar \epsilon^{ad} G
\end{equation}
respectively. The later equation is not equivalent to~\eqref{eq:Delta-comm}
while the former is in fact equivalent. It can also be rewritten
in terms of $\Delta_a^b=\epsilon_{ac}\Delta^{cb}$ as
\begin{equation}
\commut{\Delta_a^b}{\Delta^d_b}= i\hbar g_a^d G\,.
\end{equation}

\bigskip

It follows from nilpotency~\eqref{eq:nilpotency-short} and ghost number
grading~\eqref{eq:ghost} of
$\Omega$ that it can be considered as a BRST charge
of a gauge system equivalent to the original second
class one.  At the same time $\bar\Omega$ can be understood
as respective anti-BRST charge. To illustrate the idea consider
the simplest case where original second-class constraints
are ``Abelian'' i.e. their commutators form a constant
operator-valued matrix $\commut{\theta_\alpha}{\theta_\beta}
=\omega_{\alpha\beta}=\mathrm{const}_{\alpha\beta}$. In this
case one can identify constraints ${\bar\theta}^\alpha$
with conversion variables $\phi^\alpha$. The simplest choice
is to require the following commutation relations of variables
$\phi$:
\begin{equation}
\commut{\phi^\alpha}{\phi^\beta}=i\hbar\,\omega^{\alpha\beta}(-1)^{\ip{\beta}}\,,\qquad 
\omega_{\alpha\gamma}\omega^{\gamma\beta}=\delta_\alpha^\beta\,.
\end{equation}
One can then check that $\myth$ and $\bar\myth$
given by
\begin{equation}
\myth=\cC^\alpha \theta_\alpha\,, \qquad
\bar\myth= \phi^\alpha  \bP_\alpha
\end{equation}
are such that equations \eqref{eq:basic} hold. Indeed, this fact can be 
easily seen
from the explicit expressions of $\Delta,\bar\Delta$ and $Z$
\begin{equation}
\begin{gathered}
\Delta=\half \cC^\alpha \omega_{\alpha\beta}\cC^\beta\,, \qquad 
\bar\Delta=\half \bP_\alpha \omega^{\alpha\beta}\bP_\beta(-)^{\ip{\beta}}\,,\\
Z=\theta_\alpha \phi^\alpha\,.
\end{gathered}
\end{equation}
The respective BRST and anti-BRST charges $\Omega$ and $\bar\Omega$ have the 
form
\begin{equation}
\begin{aligned}
\Omega&=\myth+(i\hbar)^{-1}\commut{Z}{\myth}
=
\cC^\alpha(\theta_\alpha - \omega_{\alpha\beta}\phi^\beta)\,,
\\
\bar\Omega&=\bar\myth+(i\hbar)^{-1}\commut{Z}{\bar\myth}
=
(\phi^\alpha + \theta_\beta \omega^{\beta\alpha} 
(-1)^{\ip{\alpha}})\bP_\alpha\,.
\end{aligned}\label{eq:solution-exp}
\end{equation}
\bigskip

Let us also consider explicitly the lowest order equation
on the constraints and structure functions encoded in Eqs.~\eqref{eq:basic}.
The most general form of $\myth^a$ allowed by the ghost number
and Grassmann parity prescriptions read as:
\begin{equation}
\begin{split}
\myth&=\myth^1=\cC^\alpha \theta_\alpha
+
\half \cC^\beta \cC^\alpha \myths^\gamma_{\alpha\beta}
\bP_\gamma(-1)^{\ip{\beta}+\ip{\gamma}}\,+\,\ldots\,,
\\
\bar\myth&=\myth^2=\bar\theta^\alpha\bP_\alpha(-1)^{\ip{\alpha}}+
\half \cC^\gamma {\bar \myths}^{\alpha\beta}_\gamma
\bP_\beta \bP_\alpha (-1)^{\ip{\beta}}\,+\,\ldots\,.
\end{split}
\end{equation}
The expansion for $\Delta^{ab}$ then has the form
\begin{gather}
\Delta=- \half\cC^\beta\cC^\alpha\Delta_{\alpha\beta}(-1)^{\ip{\beta}} +\ldots\,,
\qquad 
i\hbar\,\Delta_{\alpha\beta}=\commut{\theta_\alpha}{\theta_\beta}-
i\hbar\,\myths^\gamma_{\alpha\beta}\theta_\gamma\,,
\\
\bar\Delta=
-\half \bar\Delta^{\alpha\beta}\bP_\beta \bP_\alpha(-1)^{\ip{\beta}} +\ldots \,,
\qquad 
i\hbar\,\bar\Delta^{\alpha\beta}=\commut{\bar\theta^\alpha}{\bar\theta^\beta}-
i\hbar\bar\theta^\gamma{\bar \myths}_\gamma^{\alpha\beta}\,,
\end{gather}
and
\begin{equation}
  \begin{gathered}
    Z=Z_0+\cC^\alpha Z_\alpha^\beta \bP_\beta(-1)^{\ip{\beta}}+\ldots\,, \\
    Z_0=\bar\theta^\alpha \theta_\alpha\,, \qquad
    i\hbar Z_\alpha^\beta
    =\commut{\theta_\alpha}{\bar\theta^\beta} 
       -i\hbar{\bar \myths}^{\beta \gamma}_\alpha  \theta_\gamma
       -i\hbar\bar\theta^\gamma \myths^\beta_{\gamma\alpha}-O(\hbar^2)\,.
\end{gathered}
\end{equation}
In the lowest orders in ghost variables equations~\eqref{eq:basic}
then imply
\begin{gather}
\Delta_{\alpha\beta}\bar\Delta^{\beta\gamma}(-1)^{\ip{\beta}}+O(\hbar)=\delta_\alpha^\gamma\,,\\
\commut{Z_0}{\Delta_{\alpha\beta}}
+i\hbar\,Z^\gamma_{\alpha}\Delta_{\gamma \beta}
-i\hbar\,Z^\gamma_{\beta}\Delta_{\gamma \alpha}(-1)^{\ip{\alpha}\ip{\beta}}
+O(\hbar^2)
=\,\,0
\,,\\
\commut{{\bar\Delta}^{\alpha\beta}}{Z_0}
+i\hbar{\bar\Delta}^{\alpha \gamma}  Z_\gamma^{\beta }
-i\hbar{\bar\Delta}^{\beta \gamma}  Z_\gamma^{\alpha }(-1)^{\ip{\alpha}\ip{\beta}}
+O(\hbar^2)
=\,\,0\,.
\end{gather}

\bigskip

To complete the section let us also present relations between
$\Omega^a=\myth^a+(i\hbar)^{-1}\commut{Z}{\myth^a}$,
$\myth^a$, and $\Delta^{ab}$, which are simple consequences 
of the basic equations~\eqref{eq:basic}. First, it is easy to see that
\begin{gather}
\commut{\Omega}{\bar\myth}=i\hbar(Z-G) \,, \qquad 
\commut{\bar\Omega}{\myth}=i\hbar(Z+G) \,,\\
\commut{\Omega}{\myth}=2\, i\hbar \Delta \,, \qquad 
\commut{\bar\Omega}{\bar\myth}=2\,i\hbar \bar\Delta\,,
\end{gather}
or in the short-hand notations
\begin{equation}
\commut{\Omega^a}{\myth^b}=i\hbar(2\Delta^{ab}-\epsilon^{ab}G)\,.
\end{equation}
One further arrives at
\begin{gather}
  \commut{Z}{\Omega^a}=i\hbar\Omega^a\,,\\
\commut{\Delta}{\Omega}=0\,, \qquad \commut{\bar\Delta}{\bar\Omega}=0\,,
 \end{gather}
and
\begin{gather}
  \commut{\Omega}{\bar\Delta}=i \hbar \bar\Omega\,,\qquad 
  \commut{\bar\Omega}{\Delta}=i \hbar \Omega\,.
\end{gather}

An important consequence of the above relations is the fact that operator 
algebra
generated by $\myth,\bar\myth$ is a finite-dimensional Lie superalgebra
that contains all the generating operators. As a basis in the algebra one
can take the following operators:
\begin{equation}
\myth,\bar\myth,G,\Delta,\bar\Delta,Z,\Omega,\bar\Omega\,.
\end{equation} 

\section{Hamiltonian} \label{sec:Hamiltonian}
Besides BRST generators ${\Omega}^a$ formulation of 
the quantum theory requires
specification of a unitarizing Hamiltonian. The first step
is to introduce pre-Hamiltonian $\H$ that satisfies
\begin{equation}
\commut{\Omega^a}{\H}=0\,, \qquad \p{\H}=0\,,\quad \gh{\H}=0
\label{eq:Hcond}
\end{equation}
and whose expansion with respect to ghost variables starts as
\begin{equation}
\H=H+\cC^\alpha V^\beta_\alpha \bP_\beta (-1)^{\ip{\beta}}\,+\,\ldots\,.
\end{equation}
Here, the ghost-independent term $H$ is an ``original Hamiltonian''.
In the lowest orders in ghost variables Eq.~\eqref{eq:Hcond} implies
\begin{equation}
\commut{T_\alpha}{H}= i\hbar V_\alpha^\beta T_\beta\,,
\qquad
\commut{{\bar{T}}^\alpha}{H}
= - i \hbar {\bar{T}}^\beta V_\beta^\alpha\,.
\end{equation}
Here, we made use explicit expansions~\eqref{eq:Omega-exp}
of $\Omega^1$ and $\Omega^2$ with respect to the ghost variables.

A next step in gauge fixing is to introduce nonminimal variables
$\bC_\alpha,\cP^\alpha$ and $\lambda^\alpha,\pi_\alpha$. They are assumed to 
have
the standard Grassmann parities and ghost numbers
\begin{equation}
  \begin{split}
    \p{\cP^\alpha}=\p{\bC_\alpha}=\ip{\alpha}+1\,, \qquad
    &\p{\lambda^\alpha}=\p{\pi_\alpha}=\ip{\alpha}\,,\\
    \gh{\cP^\alpha}=1 \,, \quad \gh{\bC_\alpha}=-1\,, \qquad
    &\gh{\lambda^\alpha}=0\,, \quad \gh{\pi_\alpha}=0
\end{split}
\end{equation}
and are also subjected to the following commutation relations
\begin{equation}
  \commut{\cP^\alpha}{\bC_\beta}=i\hbar\, \delta^\alpha_\beta\,, \qquad
  \commut{\lambda^\alpha}{\pi_\beta}=i\hbar\, \delta^\alpha_\beta\,.
\end{equation}

Nonminimal BRST charges are introduced according to
\begin{equation}
Q=\Omega+\pi_\alpha \cP^\alpha\,, \qquad \bar Q = \bar{\Omega}
+\bC_\alpha \lambda^\alpha\,.
\end{equation}
We use unified notation $Q^a$ with $Q^1=Q$ and $Q^2=\bar Q$. Unlike
the $sp(2)$ symmetric formalism~\cite{BLT} these nilpotent charges
do not commute:
\begin{equation}
\commut{Q}{\bar Q}=\half g_{ab}\commut{Q^a}{Q^b}=i\hbar\,
(\pi_\alpha \lambda^\alpha+ \bC_\alpha \cP^\alpha)\,.
\end{equation}

The constrained system under consideration can be described just
by nilpotent BRST charge $\Omega$ and BRST-anti-BRST-invariant Hamiltonian 
$\H$ (it can be also equivalently described by $\bar\Omega$
and $\H$). It then follows that there exist two natural proposals for
unitarizing Hamiltonian
\begin{equation}
\H_{complete}=\H+(i\hbar)^{-1}\commut{Q}{\bar\Psi}\qquad \text{and} \qquad 
\bar\H_{complete}=\H+(i\hbar)^{-1}\commut{\bar Q}{\Psi}\,,
\end{equation}
where $\Psi$ and $\bar\Psi$ are respective gauge fixing fermions.
In order to remove gauge degeneracy it is sufficient to take
$\Psi$ and $\bar\Psi$ in the ``minimal'' form
\begin{equation}
  \bar\Psi=\bC_\alpha \chi^\alpha+\bP_\alpha \lambda^\alpha\,, \qquad
  \Psi=\bar\chi_\alpha \cP^\alpha +\pi_\alpha \cC^\alpha\,,
\end{equation}
where $\chi$ and $\bar\chi$ are gauge conditions which
are supposed to be nondegenerate in the following sense:
matrices 
\begin{equation}
\pb{\chi^\alpha}{T_\beta}\,, \qquad \quad
\pb{\bar\chi_\alpha}{{\bar{T}}^\beta}
\end{equation}
are nondegenerate. One can see that $\chi$ and $\bar\chi$ are to be understood
as gauge fixing conditions associated to constraints $T$ and
$\bar{T}$ respectively.

In spite of the fact that each pair $Q,\bar \Psi$ or $\bar Q, \Psi$
can be used to describe correct physics we are interested in the
class of unitarizing Hamiltonians (and gauge fixing conditions)
respecting the symmetry between $Q$ and $\bar Q$. To this end let
us consider a class of $\H_{complete}$ corresponding to
$\bar\Psi$ of the form $\bar\Psi=(i\hbar)^{-1}\commut{\bar Q}{F}$
for some even operator $F$ of vanishing ghost number. The expression
for $\H_{complete}$ then takes the form
\begin{equation}
  \H_{complete}=\H+(i\hbar)^{-2}\commut{Q}{\commut{\bar Q}{F}}\,.
\end{equation}
It can be represented as
\begin{multline}
\H_{complete}=\H-\half (i\hbar)^{-2} \epsilon_{ab}\commut{Q^a}{\commut{Q^b}{F}}
+\frac{1}{4}(i\hbar)^{-2} g_{ab}\commut{\commut{Q^a}{Q^b}}{F}
=
\\
=
\H-\half(i\hbar)^{-2} \epsilon_{ab}\commut{Q^a}{\commut{Q^b}{F}}
+\frac{1}{2}(i\hbar)^{-1}\commut{\pi_\alpha\lambda^\alpha+\bC_\alpha 
\cP^\alpha}{F}
\,.
\end{multline}
We note that besides the third term the expression for $\H_{complete}$
coincides with that in the $sp(2)$-symmetric formalism. However,
presence of this (actually unavoidable) term reflects breaking down of
the $sp(2)$ symmetry. The residual symmetry is that preserving both
$\epsilon_{ab}$ and $g_{ab}$ and its algebra can be shown to be
isomorphic to $u(1)$.

We also note that discussion of the unitarizing Hamiltonian also
applies to any commuting and nilpotent BRST charges ${\Omega}^a$ with
appropriate ghost numbers and boundary conditions. In particular, the
same expression works for a wider class of BRST charges considered in
Sec.~\bref{sec:basics}.

\section{Conclusion} 
We have constructed a conversion scheme that results in the
BRST--anti-BRST symmetric formulation of the effective first-class
constrained (gauge) system.  This is achieved by requiring an explicit
symmetry between the original second-class constraints $\theta_\alpha$
and the extra degrees of freedom (conversion variables).  These
degrees of freedom enter the formalism as new constraints
$\bar\theta^\alpha$ dual to the original constraints $\theta_\alpha$.
In the BFV--BRST framework, the two sets of constraints are encoded in
a pair of generators $\myth$ and $\bar\myth$.  At the level of the
converted system, they are promoted to nilpotent and commuting
generators $\Omega$ and $\bar\Omega$ that are identified as BRST and
anti-BRST charges.  The symmetry between the constraints $\theta$ and
$\bar\theta$ then appears as a BRST-anti-BRST invariance of the
effective system.

It seems natural that generating equations of the BRST-anti-BRST
symmetric conversion can be recast to a master equation form if one
makes use of the quantum antibracket structure~\cite{BMQA} (see
also~\cite{BTQA}).  Indeed, a quantum antibracket master equation
encodes projector properties of the ``master action'' (either $Z$ or
$S^a_b$ in the paper), which allow constructing nilpotent generators
$\Omega$ and $\bar\Omega$ from the original generators $\myth$ and
$\bar\myth$.  {}From this general standpoint, the conversion can be
understood as a certain projection procedure.

The existence of the BRST--anti-BRST symmetric conversion follows from
the explicit solution~\eqref{eq:solution-exp} given in
Sec.~\bref{sec:special} for the ``Abelian'' constraints (in the
general situation, one can always bring constraints to an ``Abelian''
basis and then apply the inverse transformation to the generating
operators.  Therefore, there always exists a solution to the
generating equations that correctly describes the physical sector).
But there remains an open problem to describe the structure of a
general solution to the generating equations.  The difficulties are
here related to a rather nonstandard structure of the equations.

As a final remark, it is worth mentioning another point of view on the
method developed in the paper.  Instead of treating the dual
constraints $\bar\theta_\alpha$ as extra variables, we can assume that
the second-class system at hand is determined by a set of second-class
constraints that splits into two halves $\theta_\alpha$ and
$\bar\theta_\alpha$ such that each half is itself a set of
second-class constraints.  In this case, our approach results in a
BRST--anti-BRST symmetric effective gauge system that involves the
constraints $\theta$ and $\bar\theta$ in a symmetric way.  In this
approach, however, the constraints $\theta$ and $\bar\theta$ are
required to satisfy some additional relations coming from generating equations.

\subsection*{Acknowledgments} We are grateful to 
A.M.~Semikhatov and I.V.~Tyutin for illuminating discussions.  The
work of {I.A.\,Batalin} was supported by the President grant
00-15-96566, RFBR grant 02-02-16944, and INTAS grant 00-00262.  The work
of {M.A.\,Grigoriev} was supported by the RFBR grant 02-01-00930,
INTAS grant 00-00262, and RFBR grant 02-01-06096 for support of young 
scientists. MG also wishes to thank ``Actions de Recherche Concert{\'e}es'' of
the ``Direction de la Recherche Scientifique - Communaut{\'e}
Fran{\c c}aise de Belgique'', and IISN-Belgium (convention 4.4505.86).


\begin{thebibliography}{33}
\bibitem{BFV}
         Batalin I.A. and  Vilkovisky G.A.,
         {\em Relativistic S-matrix of dynamical systems with boson
         and fermion constraints,}
         Phys. Lett. {\bf B 69} (1977) 309-312.          \\
         Fradkin E.S. and Vilkovisky G.A. {\em Quantization
         of relativistic systems with
         constraints,}  Phys. Lett. {\bf B 55} (1975) 224-226.  \\
         Batalin I.A. and Fradkin E.S.,
         {\em  A generalized canonical formalism and quantization of
         reducible gauge theories,}
           Phys.Lett. {\bf B 122 }     (1983) 157-164.   \\
         Batalin I.A. and Fradkin E.S.
         {\em Operatorial quantization of relativistic dynamical systems
         subject to first class constraints,}
         Phys.Lett. {\bf B 128} 303-308 (1983).             \\
         Batalin I.A., Fradkin E.S.
         {\em Operatorial quantization of dynamical systems subject to
         constraints. A further study of the construction,}
         Ann. Inst. Henri Poincar\'e  {\bf 49} (1988) 145-214.

\bibitem{HT} M.~Henneaux and C.~Teitelboim,
\textit{Quantization of gauge systems,}
{ Princeton, USA: Univ. Pr. (1992) 520 p}.



\bibitem{FSh} L.D. Faddeev, S.L. Shatashvili, {\em Realization of the
Schwinger term in the Gauss law and the possibility of
correct quantization of a theory with anomalies},
 Phys. Lett. {\bf B 167}, 225 (1986)

\bibitem{BF} I.A.~Batalin and E.S.~Fradkin,
{\em Operator Quantization Of Dynamical Systems With Irreducible
First And Second Class Constraints,} Phys.\ Lett.\ {\bf B180}, 157 (1986).

\bibitem{BF87} I.A.~Batalin and E.S.~Fradkin, {\em
Operatorial Quantization Of Dynamical Systems Subject To Second Class
Constraints,} Nucl.\ Phys.\ {\bf B279}, 514 (1987).




\bibitem{Split}
I.~A.~Batalin, S.~L.~Lyakhovich and I.~V.~Tyutin,
{\em ``Split Involution And Second Class Constraints,''}
Mod.\ Phys.\ Lett.\ A {\bf 7}, 1931 (1992).

\bibitem{BMgauge}
I.~Batalin and R.~Marnelius,
{\em ``Gauge theory of second class constraints without extra variables,''}
Mod.\ Phys.\ Lett.\ A {\bf 16}, 1505 (2001),
\hepth{hep-th/0106087}.

\bibitem{BLM-Proj}
I.~Batalin, S.~Lyakhovich and R.~Marnelius,
\textit{``Projection operator approach to general constrained systems,''}
Phys.\ Lett.\ B {\bf 534}, 201 (2002),
\hepth{hep-th/0112175}.



\bibitem{BT} I.A.~Batalin and I.V.~Tyutin, {\em
Existence theorem for the effective gauge algebra in the generalized
        canonical formalism with Abelian conversion of second class
        constraints,} Int.\ J.\ Mod.\ Phys.\ {\bf A6}, 3255 (1991).




\bibitem{BFF}
I.A.~Batalin, E.S.~Fradkin and T.E.~Fradkina,
{\em Generalized Canonical Quantization Of Dynamical Systems With
  Constraints And Curved Phase Space,}
Nucl.\ Phys.\ {\bf B332}, 723 (1990).

\bibitem{BGL}
I.A.~Batalin, M.A.~Grigoriev and S.L.~Lyakhovich,
{\em ``Star product for second class constraint systems
from a BRST theory,''}
Theor.\ Math.\ Phys.\  {\bf 128}, 1109 (2001)
[Teor.\ Mat.\ Fiz.\  {\bf 128}, 324 (2001)],
\hepth{hep-th/0101089}.








\bibitem{BMQA}
I.~Batalin and R.~Marnelius,
\textit{``General quantum antibrackets,''}
Theor.\ Math.\ Phys.\  {\bf 120}, 1115 (1999)
[Teor.\ Mat.\ Fiz.\  {\bf 120}, 358 (1999)],
\hepth{hep-th/9905083}.
{\em ``Quantum antibrackets,''}
Phys.\ Lett.\ B {\bf 434}, 312 (1998),
\hepth{hep-th/9805084}.


\bibitem{BLT}
I.~A.~Batalin, P.~M.~Lavrov and I.~V.~Tyutin,
{\em ``Extended Brst Quantization Of Gauge Theories In The Generalized 
Canonical Formalism,''}
J.\ Math.\ Phys.\  {\bf 31}, 6 (1990).

\bibitem{BTQA}
I.A.~Batalin and I.V.~Tyutin,
{\em ``BRST-Invariant Constraint Algebra in Terms of Commutators and Quantum 
Antibrackets''}, to appear in Theor. Math. Phys.,
\hepth{hep-th/0301043}.
\end{thebibliography}
\end{document}